# A Secure Clustering Protocol with Fuzzy Trust Evaluation and Outlier Detection for Industrial Wireless Sensor Networks

Liu Yang, Yinzhi Lu, Simon X. Yang, *Senior Member*, *IEEE*, Tan Guo, and Zhifang Liang

*Abstract*—Security is one of the major concerns in Industrial Wireless Sensor Networks (IWSNs). To assure the security in clustered IWSNs, this paper presents a secure clustering protocol with fuzzy trust evaluation and outlier detection (SCFTO). Firstly, to deal with the transmission uncertainty in an open wireless medium, an interval type-2 fuzzy logic controller is adopted to estimate the trusts. And then a density based outlier detection mechanism is introduced to acquire an adaptive trust threshold used to isolate the malicious nodes from being cluster heads. Finally, a fuzzy based cluster heads election method is proposed to achieve a balance between energy saving and security assurance, so that a normal sensor node with more residual energy or less confidence on other nodes has higher probability to be the cluster head. Extensive experiments verify that our secure clustering protocol can effectively defend the network against attacks from internal malicious or compromised nodes.

## I. Introduction

RECENTLY, Industrial Wireless Sensor Networks (IWSNs) have emerged as an efficient and cost effective solution for industrial Internet of Things applications including automation, process control, monitoring, and automotive [1]. An IWSN consists of large numbers of sensor nodes which are either installed on key equipment or deployed in on-site industrial environment [2]. They can automatically organize into an ad-hoc network to perform various tasks for industrial plants. The sensor data is ultimately analyzed and correlated by a sink node to identify business events and invoke business processes that can optimize operation, maintenance, and automation in industrial production [3].

Since their deployment in valuable industrial applications and unique characteristics of self-organization, constrained resources, and dynamic topology, IWSNs are vulnerable to many attacks especially during the routing process [4], [5]. Various routing attacks may cause transmission failures and time delay while reliability and timeliness are the fundamental performance requirements of data acquisition in IWSNs [3]. Attacks can easily incur massive economic losses to industries and even threaten the security and stability of the local community [6]. Hence, it is critically important to resist security threats and assure reliable and real-time data transmission in IWSNs. Traditional authentication mechanisms can defend the network against most external attacks. However, attacks from internal malicious or compromised nodes are difficult to be detected since these nodes have authenticated identities [7]. To address this problem, trust management schemes are adopted to establish the trust relationship between sensor nodes [8]–[10]. A sensor node in the network transmits data to the trusted neighbors according to the trust management system. Then it overhears the transmissions and reports the abnormal behaviors to enhance or update the trust management system [11].

Usually, sensor nodes estimate the trust values of other nodes based on direct or indirect observations. However, defining the trust is not a crisp task, rather it can range between certain limits [12]. Then there exists uncertainty for the conversion between qualitative and quantitative trust estimations. Moreover, the workforce in industrial fields can be exposed to hazardous environments or unexpected accidents due to various interferences unfriendly to wireless communication [13]. Hence, link failures may occur accidentally that results in the inaccuracy, incompleteness, and imprecision of overheard transmissions [14]. To solve these problems, this paper presents a secure clustering protocol with fuzzy trust evaluation and outlier detection for IWSNs. The main contributions are summarized as follows:

1) We present a novel cluster based security model which consists of transmission overhearing, fuzzy trust evaluation, and outlier detection.

2) We introduce a Markov chain with two states to model the channel quality of an open wireless medium.

3) We construct an interval type-2 (IT2) fuzzy logic controller (FLC) to estimate trust values of sensor nodes.

4) We propose a density based outlier detection scheme that is used to isolate the malicious nodes from being cluster heads.

5) We present a fuzzy based clustering protocol by balancing the energy saving and security assurance.

Manuscript received April 20, 2020; revised July 17, 2020; accepted August 18, 2020. This work was supported in part by the National Natural Science Foundation of China under Grant 61801072, in part by the Chongqing Science and Technology Commission under Grant cstc2018jcyjAX0344, and in part by the National Key Research and Development Program of China under Grant 2019YFB2102001. (*Corresponding authors*: *Liu Yang; Simon X. Yang.*)

L. Yang, Y. Lu, T. Guo, and Z. Liang are with the School of Communication and Information Engineering, Chongqing University of Posts and Telecommunications, Chongqing, China (e-mail: yangliu@cqupt.edu.cn; henanluyinzhi@163.com; guot@cqupt.edu.cn; liangzf@cqupt.edu.cn).

S. X. Yang is with the Advanced Robotics and Intelligent Systems Laboratory, School of Engineering, University of Guelph, Guelph, ON N1G2W1, Canada (e-mail: syang@uoguelph.ca).

6) We verify the performance of our secure clustering protocol in terms of number of malicious clusters, number of attacks, network lifetime, and network throughput.

The rest of this paper is organized as follows: We summarize the related work in section II, and give the motivation in section III. System model and assumptions are described in section IV. In section V and VI, we present our fuzzy trust evaluation method and outlier detection scheme respectively. The detail of our secure clustering protocol is proposed in section VII, and the performance is verified in section VIII. Finally, we give the conclusions in section IX.

## II. RELATED WORK

Recently, a substantial amount of researches are performed in sensor networks where trust management is adopted to deal with the secure routing or clustering issues.

### A. Non-cluster based trust evaluation

In [15], an active detection-based security and trust routing (ActiveTrust) scheme is presented for WSNs. To assure routing security, a number of detection routes are actively created to calculate the nodal trust which is the weighted fusion of direct and recommendation trusts. To calculate the direct trust, an attenuation function is constructed that gives higher weight to the latest detected transmission. If a node is found to be malicious in the latest detection, then its direct trust is low enough so that it needs a period of cooperation to be reconsidered as a router. ActiveTrust can well defend attacks against the trust management system. However, energy is not considered during the route discovery process.

A trust and energy aware secure routing protocol (TESRP) is presented in [16] to discover and isolate misbehaving nodes. Beta distribution function is used to calculate the packet forwarding ratio of neighbor nodes based on monitored transmissions. Considering the existence of misbehaving and faulty nodes, a node is regarded as either malicious or well based on a specified packet forwarding rate threshold. Moreover, a composite routing function is designed to select the routing nodes which are trustworthy, energy-efficient, and with the shortest route to destination. TESRP is a lightweight secure routing protocol. However, its composite routing function with equal weights to trust, remaining energy, and hop counts is inefficient for secure route discovery.

In [17], an energy-aware and secure routing scheme with trust (ESRT) is presented for disaster response in WSNs. It consists of trust model, trust database, trust route discovery, and trust route maintenance. A weighted aggregation of direct trust, indirect trust, and expected positive probability is used to estimate the trust of nodes by the trust model. The necessary information needed for trust estimating is stored in trust database. The trust, remaining energy, and hop count are considered by trust route discovery system to find the energy-aware and shortest route consisting of trusted nodes. If the discovered route is no longer available or does not meet the trust and energy requirements, route maintenance will be initiated to inform the source node to find a new trusted path. The use of expected positive probability in ESRT can improve the trust estimation process since this probability is evaluated based on personal experience. However, the trust aggregation mechanism is not an adaptive one.

In [18], a trust management-based secure routing scheme (TMSRS) is presented for IWSNs with fog computing. The trust values of sensor nodes are estimated based on a Gaussian distribution. Moreover, a grey decision making is adopted to select the trusted and energy-efficient routing nodes. TMSRS can achieve a well trade-off between security, transmission performance, and energy consumption. However, the trust value of a malicious node declines slowly under the case of continuous noncooperation that may result in attacks against the trust management system.

### B. Cluster based trust evaluation

Cluster based network model is very suitable for sensor networks due to it is energy efficient and scalable. To manage the trust system in clustered sensor networks, a secure double cluster heads model (DCHM) is presented for WSNs in [19]. It consists of cluster, cluster heads, and base station modules. The clustering, cluster heads election, and trust system construction or update are performed by the cluster module. Cluster heads module is responsible for weighted outliers detection, credible data fusion, fusion results and outliers uploading. According to the trust system, two cluster heads are selected within each cluster that independently performs outlier detection, data fusion, and results uploading to base station. The dissimilarity coefficient of the two fused data for a cluster is calculated by base station. Then a feedback about the coefficient and outliers is returned to the cluster module that helps to update the trust system. DCHM can well detect the malicious nodes since two cluster heads are selected within a cluster to detect the malicious independently. However, its communication overhead is considerable due to each cluster head has to collect data from the member nodes.

To address the uncertainty problem when establishing the trust relationship between sensor nodes, a secure clustering protocol with cloud based trust evaluation method (SCCT) is presented for WSNs in [20]. Expectations of trust factors including communication, message, and energy are calculated based on beta distribution. After a period of time, enough expectations of each trust factor can be acquired and then act as drops to construct absolute trust cloud. These absolute trust clouds corresponding to the trust factors are transformed into relative ones which are weighted to be immediate trust cloud. By synthesizing immediate and recommendation trust clouds, the final trust cloud is established to determine the trust grade according to a simplified similarity calculation method. SCCT can well deal with the uncertainty of trust evidences. However, its similarity calculation method is not very effective, and the complexity is not reduced prominently.

In [21], a secured QoS aware energy efficient routing (SQEER) protocol is designed based on trust and energy modelling for enhancing the security of sensor networks while optimizing the energy utilization. For constructing the trust relationships, both direct and recommendation trusts are evaluated. Moreover, the spatial and temporal constraints are used to compute the final trust score. Then the cluster heads are selected based on QoS metrics and trust scores. To perform secure routing, path-trust, energy, and hop count are considered to get the final path. SQEER can well achieve the QoS assurance. However, the maximum recommendation trust value from the trusted nodes is used as the final recommendation that cannot reflect the real behaviors of the evaluated node.

TABLE I
COMPARATIVE ANALYSIS

| Protocols | Trust mechanism | Trust evidence | Trust estimation method | Outlier detection method | Routing method | Routing node selection method |
|---|---|---|---|---|---|---|
| ActiveTrust [15] | Weighing | Data packets, neighbors interaction | Distributed, direct, indirect | Fixed trust threshold | Flat routing | Weighing |
| TESRP [16] | Beta distribution | Data packets | Distributed, direct | Fixed trust threshold | Flat routing | Weighing |
| ESRT [17] | Beta distribution, weighing | Data packets, neighbors interaction | Distributed, direct, indirect | Fixed trust threshold | Flat routing | Weighing |
| TMSRS [18] | Gaussian distribution | Data packets, neighbors interaction | Distributed, direct, indirect | Gray theory | Flat routing | Weighing |
| DCHM [19] | Data mining | Data packets | Centralized, direct | Fixed trust threshold | Distributed clustering | Weighing |
| SCCT [20] | Cloud theory, weighing | Data packets, neighbors interaction | Distributed, direct, indirect | Fixed grade trust cloud | Distributed clustering | Weighing |
| SQEER [21] | Weighing | Data packets, neighbors interaction | Distributed, direct, indirect | Fixed trust threshold | Distributed clustering | Weighing |
| TKFCC [22] | Weighing | Data packets, neighbors interaction | Distributed, direct, indirect | Maximal trust | Centralized clustering | Weighing |
| Proposed SCFTO | Fuzzy, weighing | Data packets, neighbors interaction | Distributed, direct, indirect | Adaptive trust threshold | Distributed clustering | Fuzzy, weighing |

For trust and energy aware cluster head selection in sensor networks, a taylor kernel fuzzy C-means clustering (TKFCC) algorithm is proposed in [22]. Where clusters are formed based on a taylor kernel fuzzy C-means algorithm, and cluster heads are selected according to the fitness constraints of minimal distance, maximal trust, and maximal energy. TKFCC can achieve a well distribution of clusters. However, the weighing of fitness constraints is not very reasonable. Since the malicious nodes with more energy have the same opportunity with the normal ones with less energy to be the cluster heads.

Table I gives the comparative analysis of our proposed secure clustering protocol SCFTO with the recent trust based secure routing or clustering protocols for sensor networks.

## III. MOTIVATION

Trust management mechanism is an effective method to solve the security problem in IWSNs. To establish the trust relationship between sensor nodes, transmission overhearing is necessary for trust evidences collection. However, due to various interferences in an open wireless medium, some transmissions may be retransmitted or not overheard. Hence, uncertainty exists to the transmissions so that we cannot distinguish the malicious transmissions from missed ones. Moreover, it is hard to preset a suitable trust threshold that can be used to identify a node is whether malicious or not. Since a malicious node can enhance its trust level through cooperation, and the trust level of a normal one can be degraded due to error overhearing. To address these problems, we present a secure clustering protocol with fuzzy trust evaluation and outlier detection (SCFTO) for IWSNs. Our fuzzy trust evaluation method can effectively deal with the uncertainty of trust evidences. And adaptive trust thresholds can be acquired according to our outlier detection scheme.

## IV. SYSTEM MODEL AND ASSUMPTIONS

In this paper, we consider the clustered network model which is widely used in sensor networks due to its scalable and energy efficient characteristics. The security model, radio model, energy model, and some assumptions are given as follows:

### A. Security model

To resist the attacks from internal malicious or compromised nodes in clustered IWSNs, we present a trust based security model, as shown in Fig. 1. Usually, trust evaluation starts from transmission overhearing for trust evidences collection. It is crucial to decide which kind of information should be collected [23]. Since our proposed secure clustering protocol aims at enhancing the security of data forwarding in sensor networks, the data-plane information which reflects the changes of the network when attacks occur should be collected. The data plane is responsible for successfully data forwarding to destination within a given time slot. Three kinds of attacks regarding data packets can be launched by the malicious: tampering, delaying, and dropping. To defend against tampering attacks, data packets should be authenticated through traditional security mechanisms. Due to authentication failure, a tampered data packet is discarded which can be somehow regarded as the result of dropping attacks. Then the Data Forwarding Rate (DFR) is able to indicate the tampering attacks to some degree. In addition, dropping attacks such as selective forwarding and BlackHole are the most serious threat to data plane as they directly disrupt data forwarding. Hence, DFR is selected as one of the trust evidences in this paper. As delaying attacks become increasingly dangerous for real-time application systems, Data Forwarding Delay (DFD) which indicates the delaying attacks is considered as the second trust evidence. In clustered IWSNs, each member node collects the trust evidences of its cluster head for trust evaluation. Due to interferences in an open wireless medium, the collected trust evidences are with some uncertainty. To deal with this problem, a fuzzy trust evaluation system with an IT2 FLC is introduced to estimate the trust. In addition, an outlier detection system is developed to isolate the malicious nodes from being the cluster heads.

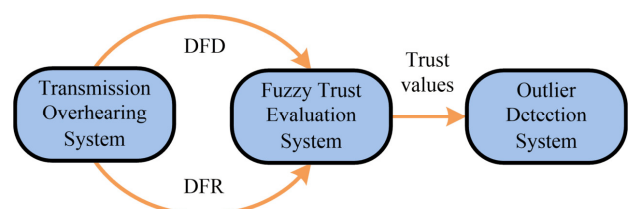

Fig. 1. Our trust based security model.

### B. Radio model

Due to the fading and interferences in an open wireless medium, the channel quality may be good or bad occasionally. Then a Markov chain with two states $S = \{s_0, s_1\}$ is introduced to model the time-varying wireless medium [24]. Here $s_0$ and $s_1$ denote the bad and good channel states respectively. The time interval $t$ that the channel quality turns to each state is a random variable that follows an exponential distribution:

$$p(t) = \begin{cases} \alpha_i \cdot e^{-\alpha_i \cdot t} & t \geq 0 \\ 0 & t < 0 \end{cases} \quad (1)$$

Where $\alpha_i$, $i \in \{0, 1\}$ are the rates of the bad and good states. Then channel state becomes bad with the probability $p_0 = \alpha_0/(\alpha_0 + \alpha_1)$, and becomes good with the probability $p_1 = \alpha_1/(\alpha_0 + \alpha_1)$.

The loss model of wireless transmission can either be free-space propagation model or two-ray ground reflection model depending on the distance $d$ between transmitter and receiver [25]. If $d$ is smaller than a distance threshold $d_0$, then the first model is considered. Otherwise, the second model is selected. Here the threshold $d_0$ can be calculated as follows:

$$d_0 = \sqrt{\varepsilon_{fs} / \varepsilon_{amp}} \quad (2)$$

Where $\varepsilon_{fs}$ and $\varepsilon_{amp}$ are the amplified characteristic constants corresponding to the above two transmission loss models.

### C. Energy model

The energy consumed by a sensor node for an $l$-bit data packet with distance $d$ can be computed as follows [21]:

$$E_{Tx}(l,d) = \begin{cases} l \cdot E_{elec} + l \cdot \varepsilon_{fs} \cdot d^2, & d < d_0 \\ l \cdot E_{elec} + l \cdot \varepsilon_{amp} \cdot d^4, & d \geq d_0 \end{cases} \quad (3)$$

Where $E_{elec}$ is the energy consumed by transmitter or receiver circuitry.

The energy spent by the receiver for an $l$-bit data packet can be calculated as follows [25]:

$$E_{Rx}(l) = l \cdot E_{elec} + l \cdot E_{DA} \quad (4)$$

Where $E_{DA}$ is the energy spent by the receiver for aggregating a one-bit data packet.

Let $D_m$ be the maximum overhearing duration for a cluster member node, and $E_m$ be the energy consumption of keeping overhearing. If a data packet is overheard by a node, an additional energy $E_h$ is consumed. Let $E_{oh}$ be the energy spent for overhearing an $l$-bit data packet. If the packet is successfully overheard after duration $d_{oh}$ ($d_{oh} \leq D_m$), $E_{oh}$ can be computed by:

$$E_{oh}(d_{oh}, l) = d_{oh} \cdot E_m + l \cdot E_h \quad (5)$$

Otherwise, the energy consumption $E_{oh}$ is calculated by $D_m \times E_m$.

### D. Assumptions

For the development of our secure clustering protocol, we make some assumptions about the sensor nodes as follows:

1) We consider an IWSN consists of large numbers of cheap sensor nodes. And because of the poor quality, these sensor nodes are likely to be compromised by the adversary.

2) We consider a distributed clustering mechanism that each node independently elects the cluster head or joins the cluster.

3) All sensor nodes are stationary or nearly stationary after deployed into the sensor field.

4) Each sensor node can monitor the transmission behaviors of neighbor nodes through overhearing. And a maximum overhearing duration is preset.

5) Three kinds of malicious nodes with heterogeneous attack abilities exist in the network.

6) Sensor nodes have homogeneous sensing, processing, and transmitting abilities. And the initial energy is homogeneous.

7) All sensor nodes can estimate the distance to transmitter based on the received signal strength, if the transmitting power is known in advance.

8) All sensor nodes can adjust its transmitting power level according to the distance to the expected destination.

## V. Fuzzy Trust Evaluation Method

IT2 FLC has been demonstrated better ability to handle uncertainty and smoother control surface than its type-1 (T1) counterpart since the membership grade of an IT2 fuzzy set (FS) is an interval [26]. The membership function of a typical IT2 FS contains an upper membership function (UMF) and a lower membership function (LMF). It maps each crisp input $z_1$ into a membership grade interval [$G_L(z_1)$, $G_U(z_1)$] that reflects the uncertainty of the grade. A wider interval means higher uncertainty to be a member of this FS. The fuzzy inference through an IT2 FLC is performed in a staged process [27], the schematic diagram is shown in Fig. 2.

To calculate trust values of sensor nodes, the trust evidences DFR and DFD are used as the inputs of an IT2 FLC. Here DFR is defined as the ratio of the amount of successfully forwarded data packet to the total amount of data packet. And DFD is defined as the ratio of the amount of delayed forwarding to the amount of successful forwarding. Three IT2 FSs corresponding to the linguistic variables *low*, *medium*, and *high* are used to describe the membership grade of DFD, as shown in Fig. 3.

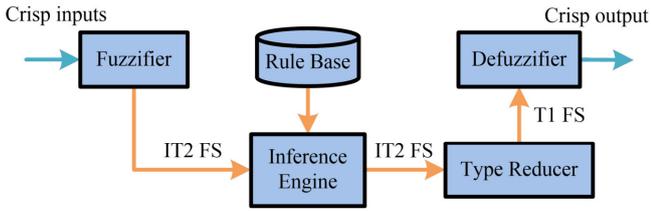

Fig. 2. The schematic diagram of a typical IT2 FLC.

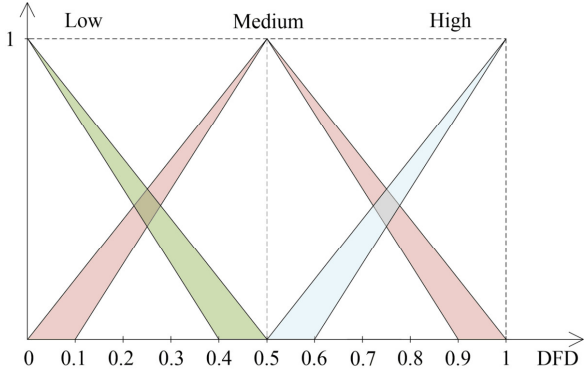

Fig. 3. IT2 FSs designed for DFD.

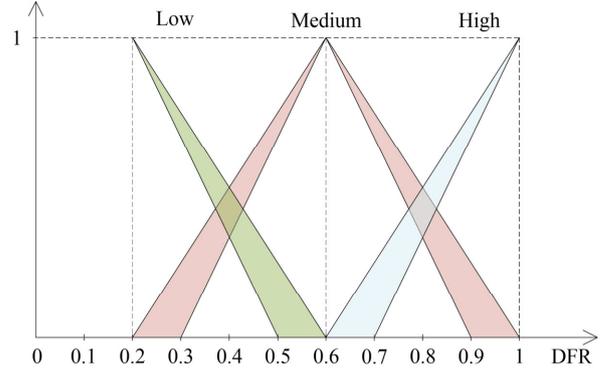

Fig. 4. IT2 FSs designed for DFR.

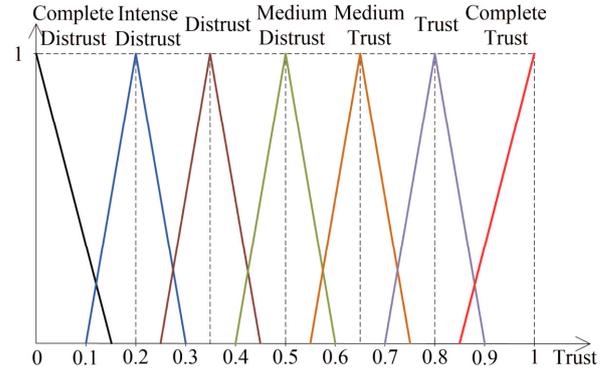

Fig. 5. T1 FSs designed for trust.

Similarly, IT2 FSs designed for DFR are shown in Fig. 4. For simplicity, seven T1 FSs with respond to the variables *complete distrust*, *intense distrust*, *distrust*, *medium distrust*, *medium trust*, *trust*, and *complete trust* are designed to assess the trust value, as shown in Fig. 5. To define the mapping of FSs between the trust evidence and trust value, ten fuzzy rules are established with *IF THEN* clauses, as shown in Table II. Especially, if DFR is smaller than 0.2, then the trust value is 0.

Let $x = (x_1, x_2)$ be the input vector of the IT2 FLC. Here $x_1$ and $x_2$ denote the values of *DFD* and *DFR*. To calculate the output trust value, six steps are performed as follows:

1) Compute the grade interval of $x_i$ ($i = 1, 2$) according to the corresponding FS for the $k^{th}$ ($k = 2, 3, ..., 10$) rule, expressed as $[G_L(x_i^k), G_U(x_i^k)]$. Where $G_L$ and $G_U$ are the lower and upper grades respectively.

2) Calculate the firing interval of $x$ for $k^{th}$ ($k = 2, 3, ..., 10$) rule, expressed as $[G_L(x^k), G_U(x^k)]$. The lower and upper firing grades can be calculated respectively as follows:

$$G_L(x^k) = G_L(x_1^k) \times G_L(x_2^k) \quad (6)$$

$$G_U(x^k) = G_U(x_1^k) \times G_U(x_2^k) \quad (7)$$

3) Calculate the output trust interval of $x$ for the $k^{th}$ ($k = 2, 3, ..., 10$) rule according to the firing interval and the corresponding trust FS, expressed as $[T_L(x^k), T_R(x^k)]$. Where $T_L$ and $T_R$ are the left and right trust values. Especially, the trust FS of the $k^{th}$ ($k = 3, 4, ..., 9$) rule is symmetrical, then two output trust intervals will be acquired if the lower firing grade is less than 1. Since the two trust intervals share the same lower and upper firing grades, values of the both firing grades will be halved in the following steps.

4) Normalize all lower and upper firing grades respectively for these fuzzy rules. In addition, all left and right trust values are sorted respectively in ascending order, expressed as $\{T_L^u, u = 1, 2, ..., l\}$ and $\{T_R^v, v = 1, 2, ..., l\}$. Corresponding to the sorted left and right trust values, the normalized firing grade intervals are $\{[G_L^u, G_U^u], u = 1, 2, ..., l\}$ and $\{[G_L^v, G_U^v], v = 1, 2, ..., l\}$. Here $l$ is an integer within [9, 16].

5) Use center-of-sets method to perform type reduction according to the normalized firing intervals and sorted trust values [25], [28]. Then a T1 FS $1/[T_L(x), T_R(x)]$ for input vector $x$ will be acquired. Here $T_L$ and $T_R$ are the final left and right trust values that can be calculated as follows:

$$T_L = \min_{m=1,2,...,l-1} \frac{\sum_{u=1}^{m} G_U^u \cdot T_L^u + \sum_{u=m+1}^{l} G_L^u \cdot T_L^u}{\sum_{u=1}^{m} G_U^u + \sum_{u=m+1}^{l} G_L^u}$$
$$= \frac{\sum_{u=1}^{M} G_U^u \cdot T_L^u + \sum_{u=M+1}^{l} G_L^u \cdot T_L^u}{\sum_{u=1}^{M} G_U^u + \sum_{u=M+1}^{l} G_L^u} \quad (8)$$

$$T_R = \min_{n=1,2,...,l-1} \frac{\sum_{v=1}^{n} G_L^v \cdot T_R^v + \sum_{v=n+1}^{l} G_U^v \cdot T_R^v}{\sum_{v=1}^{n} G_L^v + \sum_{v=n+1}^{l} G_U^v}$$
$$= \frac{\sum_{v=1}^{N} G_L^v \cdot T_R^v + \sum_{v=N+1}^{l} G_U^v \cdot T_R^v}{\sum_{v=1}^{N} G_L^v + \sum_{v=N+1}^{l} G_U^v} \quad (9)$$

TABLE II
FUZZY RULES FOR TRUST EVALUATION

| No. | DFD | DFR | Trust |
|---|---|---|---|
| 1 | × | < 0.2 | 0 |
| 2 | Low | High | Complete trust |
| 3 | Medium | High | Trust |
| 4 | High | High | Medium trust |
| 5 | Low | Medium | Medium trust |
| 6 | Medium | Medium | Medium distrust |
| 7 | High | Medium | distrust |
| 8 | Low | Low | distrust |
| 9 | Medium | Low | Intense distrust |
| 10 | High | Low | Complete distrust |

TABLE III
FUZZY RULES FOR CLUSTER HEADS ELECTION

| No. | Trust | Probability to be cluster heads ($p_{CH}$) |
|---|---|---|
| 1 | Complete trust | $p_{CH} = f_{CT}(E_{re})$ |
| 2 | Trust | $p_{CH} = f_{T}(E_{re})$ |
| 3 | Medium trust | $p_{CH} = f_{MT}(E_{re})$ |
| 4 | Medium distrust | |
| 5 | Distrust | $p_{CH} = f_{DT}(E_{re})$ |
| 6 | Intense distrust | |
| 7 | Complete distrust | |

Where $M$ is left switch point determined by $T_L^M < T_L < T_L^{M+1}$. $T_L^M, T_L^{M+1} \in \{T_L^u, u = 1, 2, ..., l\}$. $N$ is right switch point follows $T_R^N < T_R < T_R^{N+1}$. $T_R^N, T_R^{N+1} \in \{T_R^v, v = 1, 2, ..., l\}$. $T_L$ and $T_R$ can be acquired by using algorithm EIASC [29].

6) The final output trust value $T$ for the input vector $x$ can be obtained through defuzzification:

$$T(x) = \frac{T_L(x) + T_R(x)}{2} \quad (10)$$

In clustered IWSNs, a non-cluster head node usually only overhears the transmissions of its own cluster head for trust estimation. To update the trust values of sensor nodes more efficiently, trust recommendation mechanism is adopted where a non-cluster head node chooses the trusted cluster head to join cluster while requesting trust recommendations. Hence, once a node joins a cluster, it not only updates the trust value of its own cluster head through fuzzy inference. And it also updates the trust values of other nodes according to trust recommendations. If a node $s_i$ receives the recommended trust value $T(j,k)$ from its cluster head $s_j$ for node $s_k$, it updates the trust value of node $s_k$ as follows:

$$T(i,k) = \begin{cases} \dfrac{T(i,k) + T(i,j) \cdot T(j,k)}{1 + T(i,j)}, & \text{if } T(i,k) > 0 \\ T(i,j) \cdot T(j,k), & \text{otherwise} \end{cases} \quad (11)$$

Where $T(i, j)$ is the trust value of $s_j$ calculated by $s_i$ via fuzzy inference. $T(j, k)$ is the trust value of $s_k$ estimated by $s_j$ through either trust recommendation or fuzzy inference.

VI. OUTLIER DETECTION SCHEME

After the fuzzy trust evaluation, a node with more cooperation is marked with a higher trust value. However, the trust threshold used to distinguish whether a node is malicious or not is uncertain, since a malicious node can enhance its trust level through cooperation and the trust level of a normal one can be degraded as error overhearing. To deal with this problem, a density-based clustering algorithm DBSCAN [19] is adopted to develop outlier detection. Then each node analyzes the updated trust values, and concludes an adaptive trust threshold that is used to isolate the malicious from being cluster heads.

Once a node updates the trust values of other nodes through fuzzy inference or recommendations, it independently initiates an outlier detection process. The updated trust values are recorded into a set $S_T$ where each one is labeled with a number $N_{nbr}$ that represents the number of its neighbor trust values. And two trust values whose absolute difference is less than a threshold $T_{nbr}$ are neighbors. Every trust value in $S_T$ is either a Core Trust Value (CTV) or Edge Trust Value (ETV) according to its number of neighbors. If it has more neighbors than 80% of the maximum number of neighbors, then it is a CTV. In order to group the higher CTVs and their neighbor ETVs into a cluster $C_{hce}$, an iteration process with several repeated rounds is necessary. Initially, the maximum CTV in $S_T$ is moved into $C_{hce}$. In each round of the iteration process, a trust value in $S_T$ is checked whether it is a neighbor CTV or ETV of a CTV in $C_{hce}$. If the result is true, then this trust value is moved from $S_T$ to $C_{hce}$. Otherwise, it is just removed from $S_T$. The iteration process ends when $S_T$ becomes empty. Then the values in $C_{hce}$ are the representative trust levels corresponding to the normal sensor nodes in the network. Hence, the minimum value in $C_{hce}$ is considered as the trust threshold $T_{th}$ which is used to distinguish malicious nodes from normal ones.

VII. DETAIL OF OUR SECURE CLUSTERING PROTOCOL

A. Fuzzy based cluster head election

Energy and security are the major concerns for sensor nodes to elect cluster heads in IWSNs. However, energy saving and security assurance cannot be satisfied simultaneously. On the one hand, a normal node should save its energy to serve the network for a longer time. On the other hand, a normal node expects to act as the cluster head so that the malicious one has less opportunity to be the cluster head. Then a fuzzy based cluster head election scheme is presented to achieve a balance between energy saving and security assurance.

The fuzzy rules for a normal node to determine its probability of being the cluster head is given in Table III. Such as the first fuzzy rule is: If the average trust value of a certain number $N_{LCH}$ of the latest cluster heads belongs to the fuzzy set *Complete trust*, then the probability to be cluster head is calculated by the function $f_{CT}(E_{re})$. Here $E_{re}$ is the residual energy, and the probability function $f_X(E_{re})$, $X \in \{CT, T, MT, DT\}$ is given as follows:

$$f_X(E_{re}) = \left(1 - \eta_X \frac{E_{max} - E_{re}}{E_{max} - E_{min}}\right) p_X \quad (12)$$

$$s.t. \quad p_{CT} < p_T < p_{MT} < p_{DT}$$

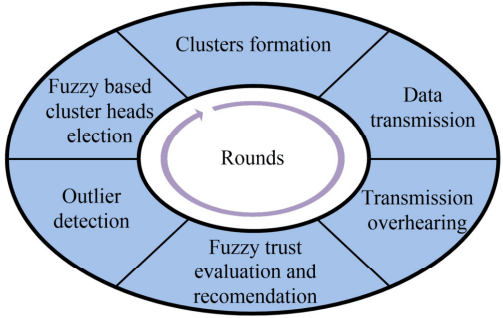

Fig. 6. Procedure of proposed protocol SCFTO.

Where $E_{max}$ is the maximum energy; $E_{min}$ is the minimum energy; $\eta_X$ is an energy factor between 0 and 1; $p_X$ is the maximum probability to be cluster heads.

Equation (12) indicates that a sensor node with more residual energy has higher probability to be the cluster head. In addition, if a sensor node has higher average trust value of its latest $N_{LCH}$ cluster heads, then it has less expect to be the cluster head since it has more confidence on the adjacent nodes. Hence, the probability to be the cluster head for each node is a balanced consideration between energy saving and security assurance.

### B. Details of our secure clustering protocol

In this section, we describe the detail of our secure clustering protocol SCFTO. Our protocol works based on rounds, the procedure is given in Fig. 6.

Initially, the probability $p_{CH}$ to be cluster heads is set to be $p_0$ for all sensor nodes. In the following rounds, each normal sensor node independently updates its own probability $p_{CH}$ according to its residual energy and the trust to other nodes. If a sensor node has not been a cluster head during the latest $1/p_{CH}$ rounds, then it decides whether to be the head in the current round $r$ according to a threshold $Th_{CH}$:

$$Th_{CH} = p_{CH} / (1 - p_{CH} \cdot (r \bmod (1/p_{CH}))) \quad (13)$$

Sensor nodes self-decided to be the heads broadcast election messages within the whole network. Other non-cluster head nodes choose their own heads to join clusters based on the distances and trusts to these heads. For any non-cluster head node, it chooses a certain number $N_{NCH}$ of the nearest cluster heads as the candidates. Usually, the trust threshold $T_{th}$ acquired through outlier detection is not stable due to inadequate direct interactions. To accelerate the convergence of the outlier detection process, the candidate cluster heads with zero trust values are firstly selected, and the closer one is given a higher priority. Otherwise, the candidate with the highest trust value is chosen as the cluster head. If the absolute difference of the trust thresholds obtained in adjacent rounds is less than a threshold $Th_d$, and this situation is maintained continuously for a certain number $N_s$ of rounds, then the outlier detection process converges. After that, each candidate head can be identified to be trusted or not according to the trust threshold $T_{th}$. Then the nearest trusted candidate is selected as the cluster head. If no trusted candidate exists, the cluster head can be selected from the candidates with zero trust value. If a sensor node does not select a proper cluster head, it has to self-declare to be the cluster head if it is eligible. Once a sensor node finds a proper cluster head, it sends a request message with its ID and residual energy. After the request time is over, each cluster head divides the time slots based on the number of requesters to avoid transmission interferences. And then it sends an acceptance message including the ID, arranged time slot, maximum energy, minimum energy, and trust recommendations to each requester.

After the constructing of clusters in one round, each cluster member node transmits its data to its own head, and followed with the transmission overhearing. Based on the overheard trust evidences, the member node estimates the trust of its cluster head through fuzzy inference, and updates the trusts to other nodes based on recommendations from the head. By analyzing the updated trust values, outlier detection is developed to update the trust threshold $T_{th}$ that used to isolate the malicious nodes from being the cluster heads in the next round.

## VIII. PERFORMANCE EVALUATION

### A. Experimental setup

Since a Markov chain with two states is used to model the channel quality, we assume the duration of a round is short enough so that the channel state does not change within a round. We call the three kinds of malicious nodes as generic, advanced, and super ones. And they account for 30, 40, and 30 percent of the malicious. We assume the three kinds of malicious sensor nodes launch dropping attacks with the probabilities $P_{sf}$, $2 \times P_{sf}$, and $3 \times P_{sf}$ respectively. And launch delaying attacks with the probabilities $P_{df}$, $2 \times P_{df}$, and $3 \times P_{df}$ respectively. Usually, a normal cluster head forwards the data packets timely for its member nodes. However, the forwarded packet may be lost due to the interferences when channel state is bad. Then the normal head has to forward the lost packet again so that a delayed forwarding event may be captured with the probability $P_{cd}$. We assume that no error overhearing event occurs if the channel quality is good. Otherwise, a forwarded packet may not be overheard with the probability $P_{no}$. A normal cluster head forwards the lost packet again with an interval of $0.5 \times D_m$. Here $D_m$ is the preset maximum overhearing duration. If the overhearing node does not overhear the packet until timeout, it adds a record of dropping attack. If a malicious cluster head intends to launch a delaying attack, it forwards the packet with a random delay no longer than $D_m$.

TABLE IV
PARAMETERS SETTING

| Parameter | value | Parameter | value |
|---|---|---|---|
| Packet size (bits) | 3000 | $D_m$ (s) | 10 |
| Control packet size (bits) | 300 | $\alpha_0$ | 3 |
| Initial energy $E_0$ (J) | 1.5 | $\alpha_1$ | 7 |
| $E_{elec}$ (nJ/bit) | 50 | $Th_d$ | 0.05 |
| $\varepsilon_{amp}$ (pJ/bit/m$^4$) | 0.0013 | $N_s$ | 60 |
| $\varepsilon_{fs}$ (pJ/bit/m$^2$) | 10 | $P_{sf}$ | 0.1 |
| $E_{DA}$ (nJ/bit/message) | 5 | $P_{df}$ | 0.1 |
| $E_h$ (nJ/bit) | 5 | $P_{cd}$ | 0.2 |
| $E_m$ (nJ/s) | 10 | $P_{no}$ | 0.2 |
| $p_0$ | 0.07 | $\eta_X$ | 0.4 |
| Initial trust value | 0 | $p_{CT}$ | 0.08 |
| $T_{nbr}$ | 0.01 | $p_T$ | 0.1 |
| $N_{LCH}$ | 10 | $p_{MT}$ | 0.12 |
| $N_{NCH}$ | 2 | $p_{DT}$ | 0.14 |

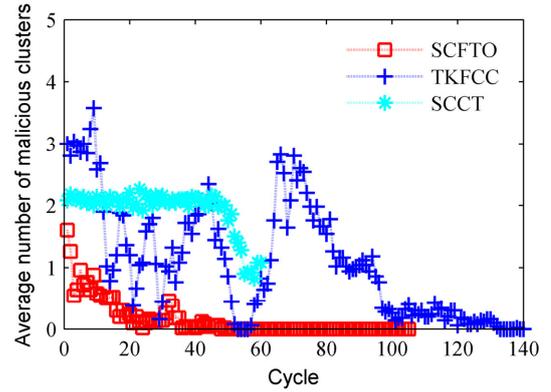

Fig. 7. Average number of malicious clusters in every cycle.

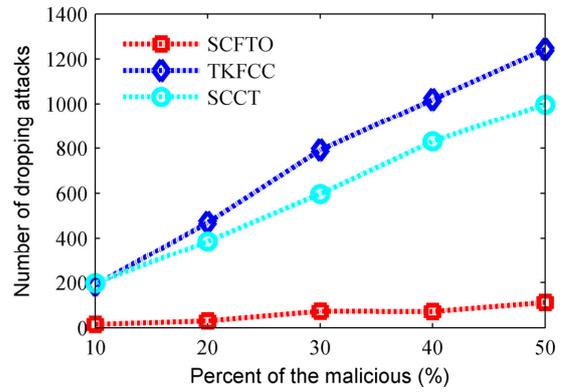

Fig. 8. Total number of dropping attacks.

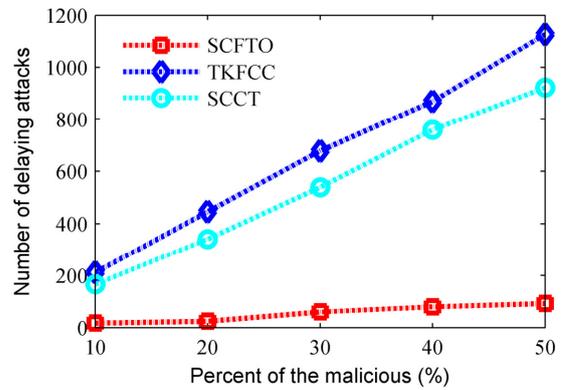

Fig. 9. Total number of delaying attacks.

Since uncertainty of trust evidences is addressed in SCCT, and a balance between security and energy is considered in TKFCC, the performance of our proposed protocol SCFTO is verified by comparing it with that of SCCT and TKFCC. In SCCT, DFD and DFR are used as the trust evaluation factors. To transform the absolute trust clouds into relative ones, the failure tolerance is set to be 0.2. To acquire the final trust cloud, the immediate trust cloud and recommendation trust cloud are synthesized with the time sensitive factor 0.6. To assess the proximity to standard grade trust clouds, totally 20 drops are randomly selected from the final trust cloud. The number of clusters in TKFCC is set to be 5 percent of total number of sensor nodes. In addition, other parameters used throughout our experiments are given in Table IV.

*B. Experiment results*

In this section, we first verify the performance of our proposed protocol SCFTO under the case that 100 sensor nodes are randomly deployed into a square sensor field with the size of 100×100 m$^2$, and the base station is placed at the coordinate (150, 50) with respect to the bottom left corner of the sensor field. And then, to further verify the scalability of our proposed protocol, we perform the comparison experiments under the network scenario of different positions of base station, network sizes, and node densities.

The comparison of average number of malicious clusters in every cycle among SCFTO, TKFCC, and SCCT is given in Fig. 7. The duration of a cycle is set to be 50 rounds, and 30 percent of malicious nodes exist in the network. This figure shows that our protocol SCFTO has fewer malicious clusters than TKFCC and SCCT in most cycles. In addition, the number of malicious clusters of SCFTO decreases with fluctuations and then reaches the minimum value 0. This case indicates that our trust based secure mechanism can effectively isolate the malicious nodes from being cluster heads after a period of trust evidences collection. The number of malicious clusters in TKFCC decreases first and then increases with fluctuations, and finally reaches the minimum. This is because the weighing of fitness constraints used in TKFCC is not adaptive, so that the malicious nodes have the chance to be cluster heads after the normal ones exhaust much energy.

With the increase of percent of malicious nodes in the network, the total number of dropping or delaying attacks increases for all these protocols, as shown in Figs. 8 and 9. In addition, SCFTO has the fewest attacks while TKFCC has the most for each percent of malicious nodes. This is because fuzziness of trust evidences is considered in both SCFTO and SCCT. Moreover, an adaptive outlier detection method and the balance mechanism between energy and security are introduced in SCFTO. Although the energy saving and security assurance are balanced in TKFCC, the fixed weighing of fitness constraints results in many attacks after most normal nodes exhaust much energy.

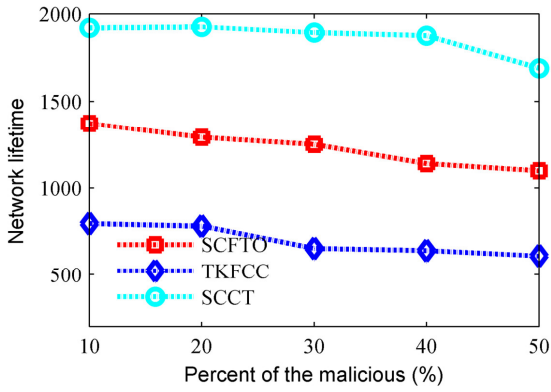

Fig. 10. Network lifetime.

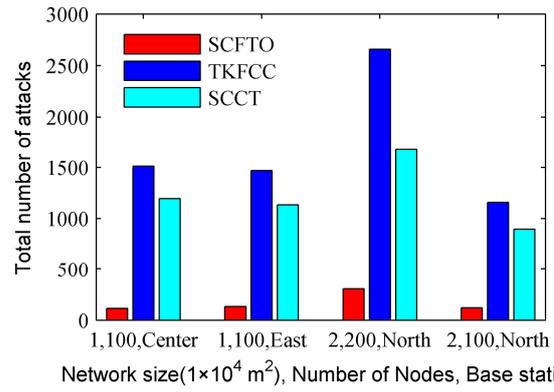

Fig. 13. Total number of attacks of the networks with different positions of base station, sizes, and node densities.

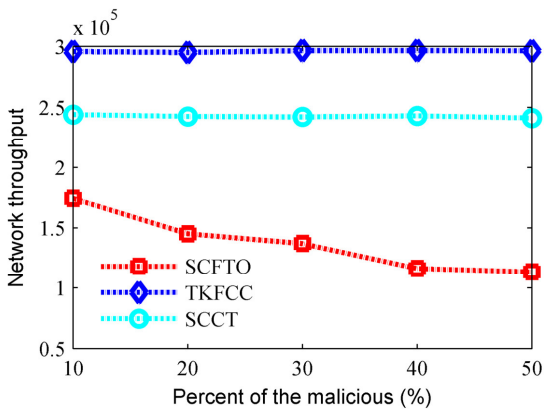

Fig. 11. Network throughput.

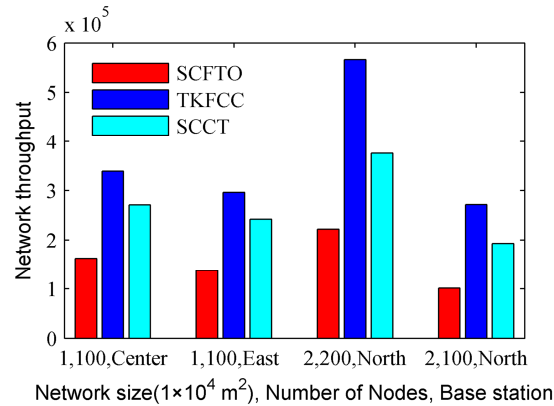

Fig. 14. Throughput of the networks with different positions of base station, sizes, and node densities.

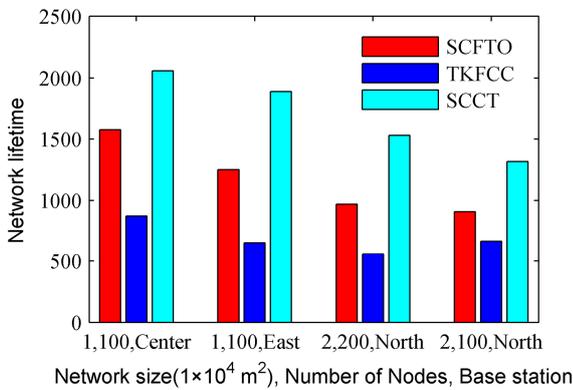

Fig. 12. Lifetime of the networks with different positions of base station, sizes, and node densities.

Fig. 10 gives the comparison of network lifetime among SCFTO, TKFCC, and SCCT for different number of malicious nodes. It shows that the network lifetime decreases with the increase of the percent of malicious nodes. This is because some malicious sensor nodes are excluded from being the cluster heads due to trust management system, so that the normal ones have to burden more data forwarding tasks when there are more malicious sensor nodes in the network. Since much more malicious nodes are excluded from being cluster heads, SCFTO has shorter network lifetime than SCCF. Although more malicious nodes are successfully elected as cluster heads to burden data transmission tasks, TKFCC has shorter network lifetime than SCFTO. This is because the energy consumption between sensor nodes is not well balanced due to the fixed weighing of the fitness constraints in TKFCC.

Fig. 11 gives the comparison of network throughput among SCFTO, TKFCC, and SCCT. With the increase of the percent of the malicious, network throughput decreases in SCFTO but almost keeps stable in TKFCC and SCCT. In addition, SCFTO has the fewest network throughputs, and TKFCC has the most. This is because malicious nodes have the least chance to be cluster heads and then contribute to data transmission in SCFTO. However, the malicious nodes have much chance to be cluster heads once the normal ones exhaust much energy in TKFCC.

The comparison of network lifetime among SCFTO, TKFCC, and SCCT for different positions of base station, network sizes, and node densities is shown in Fig. 12. For a network with 100 sensor nodes including 30 percent of the malicious deployed into the 100 × 100 m² sensor field, if the base station is placed at the center rather than on the east, all the three protocols can acquire a longer network lifetime due to the shorter average distance between sensor nodes and base station. If the network size increases to 200 × 100 m² while the number of sensor nodes is 200 and the base station is placed on the north, the network lifetime of the three protocols decreases as the longer average distance from nodes to based station. If the node density decreases from 0.01 to 0.005, network lifetime of SCFTO and SCCT decreases while that of TKFCC increases. This is because the average distance between sensor nodes increases when node density decreases. However, the energy consumption between sensor nodes can be balanced more easily in TKFCC for the network with a smaller node density.

Fig. 13 gives the comparison of total number of dropping and delaying attacks among SCFTO, TKFCC, and SCCT. This figure shows that SCFTO has fewer attacks than TKFCC and SCCT. In addition, the total number of attacks is almost the same in SCFTO for the networks with different positions of base station, sizes, and node densities while the number of sensor nodes remains constant. This is because all malicious nodes are excluded from being cluster heads once enough trust evidences are collected. If the number of sensor nodes increases from 100 to 200, the number of attacks increases for all the three protocols due to the existence of more malicious nodes. And if the node density decreases from 0.01 to 0.005, the number of attacks decreases for both TKFCC and SCCT as fewer malicious nodes exist in the adjacent area.

The comparison of network throughput among SCFTO, TKFCC, and SCCT for different positions of base station, network sizes, and node densities is shown in Fig. 14. This figure shows that when the number of sensor nodes increases from 100 to 200, the network throughput increases for all the three protocols as more sensor nodes contribute to the data collection and transmission. If the node density decreases from 0.01 to 0.005, the three protocols acquire a smaller network throughput due to the average distance between sensor nodes increases. In addition, when the base station is placed at the center instead of on the east of the sensor field, these protocols can achieve a bigger network throughput since the average distance between sensor nodes and base station decreases.

## IX. CONCLUSION

In this paper, we first present a novel trust based secure mechanism which consists of transmission overhearing, fuzzy trust evaluation, and outlier detection for IWSNs. Our secure mechanism is easy to implement and fully distributed. It can deal with the transmission uncertainty caused by interferences in an open wireless medium. And an adaptive trust threshold can be acquired to isolate the malicious from being cluster heads. And then a fuzzy based secure clustering protocol is proposed to achieve a balance between energy saving and security assurance. Simulation results show that our secure clustering protocol has better performance on defending against internal attacks than those based on cloud model and weighing mechanism. In our future work, we will focus on efficient transmission overhearing methods that can ease transmission uncertainty in an open wireless medium. Moreover, to accelerate the convergence of outlier detection process, we will further investigate effective and lightweight data preparation mechanisms to deal with the original trust evidences.

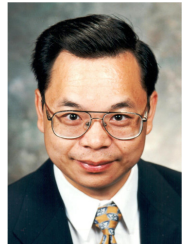
**Simon X. Yang** (S'97–M'99–SM'08) received the B.Sc. degree in engineering physics from Beijing University, Beijing, China, in 1987, the first of two M.Sc. degrees in biophysics from the Chinese Academy of Sciences, Beijing, China, in 1990, the second M.Sc. degree in electrical engineering from the University of Houston, Houston, TX, in 1996, and the Ph.D. degree in electrical and computer engineering from the University of Alberta, Edmonton, AB, Canada, in 1999.

Dr. Yang is currently a Professor and the Head of the Advanced Robotics and Intelligent Systems Laboratory at the University of Guelph, Guelph, ON, Canada. His research interests include robotics, intelligent systems, sensors and multi-sensor fusion, wireless sensor networks, control systems, machine learning, fuzzy systems, and computational neuroscience.

Prof. Yang has been very active in professional activities. He serves as the Editor-in-Chief of International Journal of Robotics and Automation, and an Associate Editor of IEEE Transactions on Cybernetics, IEEE Transactions on Artificial Intelligence, and several other journals. He has involved in the organization of many international conferences.

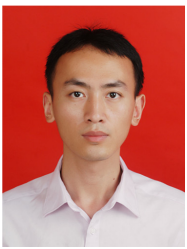
**Liu Yang** received his B.S. degree in Electronic Information Science and Technology from Qingdao University of Technology, Shandong, China, in 2010, and Ph.D. degree in Communication and Information Systems at the School of Communication Engineering, Chongqing University, Chongqing, China, in 2016. He is now a lecturer in Chongqing University of Posts and Telecommunications. His research interests include Internet of Things, data analysis, and artificial intelligence.

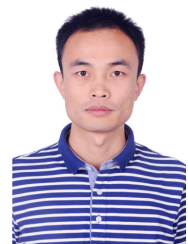
**Tan Guo** received his M.S. degree in Signal and Information Processing from Chongqing University, Chongqing, China, in 2014, and Ph.D. degree in Communication and Information Systems from Chongqing University, Chongqing, China, in 2017. He is now a lecturer in Chongqing University of Posts and Telecommunications. His research interests include Internet of Things, biometrics, pattern recognition, and machine learning.

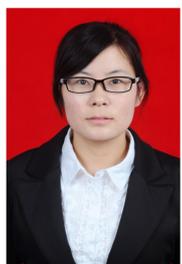
**Yinzhi Lu** received her M.S. degree in Communication and Information Systems from Chongqing University, Chongqing, China, in 2014. She is currently pursuing the Ph. D. degree in Information and Communication Engineering with the School of Communication and Information Engineering, Chongqing University of Posts and Telecommunications. She was a teaching assistant with the School of Electronic Information Engineering, Yangtze Normal University from 2014 to 2019. Her current research interests include Internet of Things, time sensitive network, and artificial intelligence.

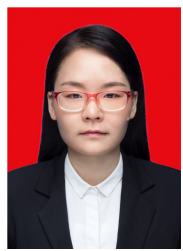
**Zhifang Liang** received her B.S. degree in Computer Science and Technology from Shanxi Normal University, Linfen, China, in 2013, and received her Ph.D. degree in intelligent signal processing from the School of Communication Engineering, Chongqing University, Chongqing, China, in 2017. Currently she works in the school of Communication and Information Engineering, Chongqing University of Posts and Telecommunications. Her research interests include electronic nose technology, Internet of Things, and machine learning.